\documentclass[draftcls,11pt,onecolumn,letterpaper]{IEEEtran}
\IEEEoverridecommandlockouts
\usepackage{subcaption}
\usepackage{enumerate}
\newcommand{\argmin}{\operatornamewithlimits{argmin}}
\usepackage{float}
\usepackage{color}
\usepackage{graphicx}
\usepackage{caption}
\usepackage{amsmath} 
\usepackage{amssymb,algorithm}  

\title{ Social Learning in a Human Society: An Experimental Study}

\author{Maziyar Hamdi, Grayden Solman, and Alan Kingstone, Vikram Krishnamurthy \thanks{Maziyar Hamdi
and  Vikram Krishnamurthy are with the Department of Electrical and Computer
Engineering and Grayden Solman and Alan Kinsgstone are with the Department of Psychology, University of British Columbia, Vancouver, Canada. }
}

\newcommand{\Y}{\mathbb{Y}}
\newcommand{\X}{\mathbb{X}}

\newcommand{\beq}{\begin{equation}}
\newcommand{\eeq}{\end{equation}}

  \def\1{{\mathbf 1}}

\begin{document}

\maketitle
\thispagestyle{empty}
\pagestyle{empty}

\begin{abstract}

\hspace{1mm} This paper presents an experimental study to investigate the learning and decision making behavior of individuals in a human society.  Social learning is used as the mathematical basis for modelling interaction of individuals that aim to perform a perceptual task interactively. A psychology experiment was conducted on a group of undergraduate students at the University of British Columbia to examine whether the decision (action) of one individual affects the decision of the subsequent individuals. The major experimental observation that stands out here is that the participants of the experiment (agents) were affected by  decisions of their partners in a relatively large fraction (60\%) of trials.  We fit a social learning model that mimics the interactions  between participants of the psychology experiment.  Mis-information propagation (also known as data incest) within the society under study is further investigated in this paper.

\end{abstract}

\begin{keywords}
Bayesian models, data incest, herding, mis-information propagation, social learning, social networks.
\end{keywords}
\section{Introduction}\label{sec:intro}

Social learning is used as a mathematical framework  to model the interaction between individuals that aim to estimate an unknown parameter  (state of the nature). Social learning in multi-agent systems seeks to investigate how decisions made by agents affect decisions made by subsequent agents. In social learning, each agent chooses its action by optimizing its local utility function. Subsequent agents then use their private observations together with the decisions of previous agents to estimate (learn) the underlying state of nature. In the last decade, social learning has been used widely in economics, marketing, political science, and sociology to model the behavior of financial markets, crowds, social groups, and social networks; see~\cite{ADLO08,AO11,Ban92,BHW92,Cha04,LADO07} and numerous references therein. Related models have been studied in the context of sequential decision making in information theory~\cite{CH70,HC70} and statistical signal processing~\cite{CSL13,KP13}.

It is perhaps surprising, then, that the behaviour of this model has yet to be tested in real human agents. This paper provides an experimental study to investigate how social learning unfolds between real human agents.  Section~\ref{sec:model} presents a  brief description of the classical Bayesian social learning model. 
 In social learning~\cite{Cha04}, agents estimate the underlying state of nature not only from their local measurements, but also from the actions of previous agents. (These previous actions  were taken by agents in response to their local measurements; therefore, these actions convey information about the underlying state). As we will describe in Section~\ref{sec:model}, the state estimation update in social learning has a drastically different structure compared to the standard optimal filtering recursion and can result in unusual behavior. In the classical social learning model,  each agent  acts once in a pre-determined order. However, in real social settings, individuals may act several times and, thus, social learning can take place with information exchange over a loopy graph  (where the agents form the vertices of the graph).
Due to the loops in the information exchange graph, {\em data incest} (mis-information) can propagate as we explain in Section~\ref{sec:model}. To study the learning behavior of individuals in a human society, we conducted a psychology experiment.  A detailed description of the psychology experiment is provided in Section~\ref{sec:setup}. The main results of our experimental study are presented in Section~\ref{sec:results}.  A social learning model is fitted to the observed interactions between real agents  performing a perceptual task interactively. We further investigate  mis-information propagation (data incest) within the society under study.

\section{Social Learning Model}\label{sec:model}
Consider  $S$  agents  performing social learning to estimate an unknown parameter (state of nature), where $S$ is a positive integer. Let $x \in \mathbb{X} = \{1, 2,\cdots, X\}$ represent the state of nature  with known prior distribution $\pi_0$, where $X$ is a positive integer number. Each agent acts in a sequential order indexed by $k = 1,2,3,\ldots$. Here, $k$ depict epochs at which events occur. These events comprise of taking observations, evaluating beliefs and choosing actions as  described below. The index $k$ depicts the historical order of events and not necessarily absolute time. However, for simplicity, we refer to $k$ as ``time" in the rest of the paper.

The agents use the following Bayesian social learning protocol to estimate the state of nature:
\paragraph*{Step 1. Private observations} To estimate the state of nature $x$, each agent records its $M$-dimensional private observation vector. At each time $k = 1, 2, 3, \ldots$, each agent $s$ ($1 \leq s \leq S$) obtains a noisy private observation $y_{[s,k]}$ from the finite set of possible observations  $\mathbb{Y} = \{1, 2,\ldots,Y\}$ with conditional probability
\begin{equation}\label{eq:B}
B_{ij} = p(y_{[ s,k]} =  j|x =  i).
\end{equation} Here, $p(\chi)$ is the probability of event $\chi$. It is assumed that the observations  $y_{[s,k]}$ are independent random variables with respect to agent $s$ and time $k$.
\paragraph*{Step 2. Private belief}Using private observations, each agent computes its private belief as \begin{equation}\label{eq:privateb1}
\mu_{[ s,k ]} = (\mu_{[ s,k ]}(i), 1\leq i\leq X),\text{where}\quad  \mu_{[ s,k ]}(i) = p\left(x =i|\Theta_{[s,k]}, y_{[s,k]}\right).
\end{equation}
where $\Theta_{[ s,k ]}$ denotes the set of actions of other agents at previous time instants that are available to agent $s$ at time $k$. Note that the agent's private belief (private opinion) is  not available to the other agents.

\paragraph*{Step 3. Local action} Based on its private belief $\mu_{[ s,k]}$, agent $s$ at time $k$ chooses an action $a_{[s,k]}$ from a finite set $ \mathbb{A} = \{1,2,\ldots,A\}$ to minimize its expected cost function. That is
\begin{equation}\label{eq:action}
a_{[s,k]} = \argmin_{a\in \mathbb{A}}\mathbf{E}\{C(x,a)|\mu_{[s,k]}\}.
\end{equation}
Here $\mathbf{E}$ denotes expectation and  $C(x,a)$ denotes the cost incurred by the agent if action $a$ is chosen when the state of nature is $x$. Note that although private observations are independent, but the actions of each agent depend on the earlier actions of all agents.

From a statistical signal processing point of view, estimating the state of nature $x$ using the above  protocol is non-standard in two ways: First, the social learning can result in  multiple rational agents taking the same action independently of their observations. A well known result in the economics literature~\cite{BHW92,Cha04}, states that if agents follow the above social learning protocol, then   after some finite time $\bar{k}$, a {\em herd of agents} occurs.
A  herd of agents takes place at time $\bar{k}$, if the actions of all agents after time $\bar{k}$ are identical, i.e., $a_{[s,k]} = a_{[s,\bar{k}]}$ for all
time  $k > \bar{k}$ and $s = 1\ldots S$. Second, (and this effect is more complex), an agent might be influenced by his own action leading to data incest (mis-information propagation). The nature of data incest can be illustrated with the following example:
Suppose an agent wrote  a poor rating of a restaurant on a social media site.  Another agent is influenced by this rating, visits the restaurant, and then also gives  a poor rating on the social media site. The first agent visits the social media site  and notices that another agent has also given the restaurant a poor rating---this double confirms her rating and she reinforces her prior belief. Data incest (mis-information propagation) in a network is defined as defined as occurring if agent $s$ naively includes the actions of other agents in the formation of her belief, when these other agents had themselves been influenced by her own actions at an earlier time.

In this paper, we fit a social learning model (find the best-fit values for $B$, $C$, and $\pi_0$) to mimic the learning behavior of real human agents. The presence and possible influence of mis-information propagation (data incest) is also investigated.

\section{Experiment Setup}\label{sec:setup}
Here, a detailed description of the psychology experiment we carried out to study the learning behavior of individuals in a human society is presented:
\begin{itemize}
\item \textit{Experiment Date:} The psychology experiment was conducted in September and October 2013.

\item \textit{Society under study:} The participants were 36 undergraduate students (18 pairs) from the department of Psychology at the University of British Columbia who completed the experiment for course credit.

\item \textit{Experiment Setup:}
Participants were asked to perform a perceptual task interactively. For each trial, two arrays of circles were given to each pair of participants, and they were asked to judge which array had the larger average diameter; that is, picking their actions. The two $4\times4$ grids of circles were generated by randomly drawing from the radii: $\{ 20, 24, 29, 35, 42\}$ (in pixels). The average diameter of each grid was computed, and if the means differed by more than 8\% or less than 4\%, new grids were made, i.e., each trial had arrays of circles differing in the average diameter length by 4-8\%\footnote{These numbers were chosen for high task-difficultly, targeting a single-exposure performance level near chance based on the results of Treisman et. al.~\cite{TG80}. }.
 One participant was chosen randomly to start each trial by choosing an action according to his observation. Thereafter, the other participant saw their partner's previous response (action) and stimulus array prior to making their own judgment; this is social learning. The participants continued choosing actions until their responses stabilized for a run of at least three (two participants did not necessarily agree, but each was fixed in her responses). In this experimental study, each participant chose an action in $\mathbb{A} = \{0,1\}$;  $a = 0$  when she judged that the left array of circles had the larger diameter and $a = 1$ when her judgments  was that the right array of circles had the larger diameter. In each trial, judgments (actions) of participants were recorded along with the amount of time taken to make that judgment.  Each pair completed a number of trials within a single 45 minute experimental session. Since trial lengths varied on the basis of response stabilization rates, the number of trials varied from pair to pair. Fig.~\ref{Fig:Samplepath} shows the judgments of two participants in a pair at different steps in one trial. In this trial, the average diameter of the left array of circles was $32.1875$ and the right array was	$30.5625$ (in pixels).

\begin{figure}[h]
\centerline{
\includegraphics[width=\textwidth]{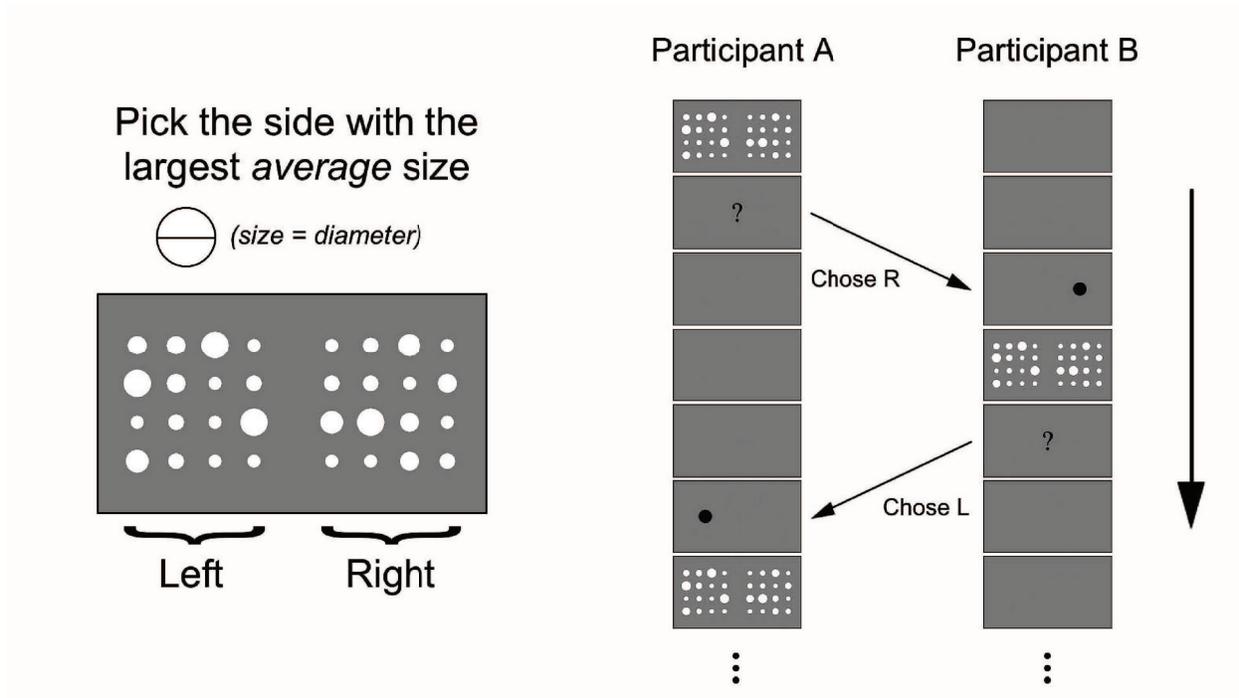}}
\caption{Two arrays of circles were given to each pair of participants on a screen. Their task is to interactively determine which side (either left or right) had the larger average diameter. The partner's previous decision was displayed on screen prior to the stimulus.}
\label{Fig:SocialSensor}
\end{figure}

 \begin{figure}[htb]
\centerline{
\includegraphics[width=.8\textwidth]{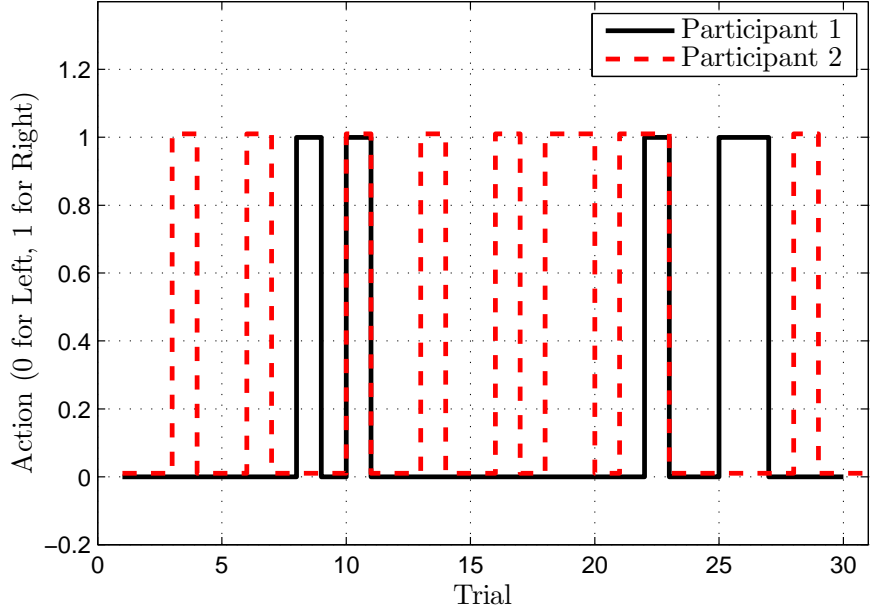}}
\caption{Actions of two participants in a group at different trials in one experiment. }
\label{Fig:Samplepath}
\end{figure}
\end{itemize}
\section{Experimental Results}\label{sec:results}
 The results of our experimental study, which are summarized in Fig.\ref{summary}, are as follows:
\begin{itemize}
\item {\em Social learning Model:} As mentioned above, each trial for a pair of participants was continued until both participants' responses stabilized. We first ask, then: {\em In what percentage of trials is an agreement made between the participants?} This percentage characterizes the degree of “herding” that occurred in the experiment. In other words, it reveals the extent to which participants’ decisions were influenced by the behaviour of their partners (i.e., social learning occurred). Interestingly, our experimental study shows that in 66\% of all trials (1102 among 1658), participants reached an agreement; that is herding occurred. Further, our experimental studies show that in 32\% of all trials, the social learning was successful and both participants made the right judgment after few interactions. To find a proper social learning model, we focus on the experiments where both participants reached an agreement. Define the social learning (SL) success rate as
    \begin{equation}\nonumber
    \text{SL Success Rate:  } \frac{\text{No. of experiments where both participants chose the correct side}}{\text{No. of experiments where both participants reached an agreement}}\cdot
    \end{equation}
   In this experimental study, the state space is $\X = \{0,1\}$ where $x = 0$, when the left array of circles has the larger diameter and $x = 1$, when the right array has the larger diameter. The initial belief for both participants is considered to be $\pi_0 = [0.5, 0.5]$. The observation state is assumed to be $\Y = \{0,1\}$. We fit a social learning model to our experimental data which gives the same success rate as the experimental study (SL Success Rate = $50\%$). The social learning parameters (probability observations, $B_{iy}= p(y_k = y|x = i), i\in \X, y \in \Y$ and the cost function $C(i,a), i \in \X, a \in \mathbb{A} $), obtained by exhaustive search, are as follows:
   \begin{align}
   &B_{iy}  = \begin{bmatrix} 0.61 & 0.39 \\ 0.41 & 0.59 \end{bmatrix}, \nonumber\\
   &C(i,a) = \begin{bmatrix} 0 & 2 \\ 2 & 0 \end{bmatrix}\nonumber.
   \end{align}

\item {\em Data incest:} Here, we study the effect of data incest on the judgments of participants in our experimental study. Since we do not have access to the private observations of individuals (almost no one has such information!), we cannot exactly verify that whether data incest changed the judgment of an individual in each step of a trial. However, two scenarios which are depicted in Fig.\ref{exp:dataincest} are used to find data incest events in the experiments. In these two events, as can be seen in Fig.\ref{exp:dataincest}, the action of the first participant at time $k$ influences the action of the second participant at time $k+1$, and thus, is double counted by the first participant at time $k+2$. As discussed above, since we do not have access to the private observations of participant, we cannot exactly say that data incest affects the action of the first participant at time $k+2$ or not. However, it is clear that in these events, data incest occurs. As we expect from the communication topology between partners, data incest occurred in relatively large percentage of trials in the experiment. More precisely, in 79\% of trials one of the data incest events shown in Fig.\ref{exp:dataincest} occurred (1303 trials with data incest among 1658 trials). Our experimental study further shows that in 21\% of experiments, data incest  resulted in the changing the decision of one of the the participants in the group, i.e., the judgment of participant at time $k+1$  differed from her judgments at time $k+2$ and $k$ in the events shown in Fig.\ref{exp:dataincest}. This experimental study reveals that data incest is quite common in social learning in human societies (happened frequently in our simple social learning setup with small number of agents) and, therefore, social learning protocols require a careful design to handle and mitigate data incest.
\begin{figure}[h]
\centering
\begin{minipage}[b]{.45\textwidth}
\hspace{-.2cm}\scalebox{.57}{\includegraphics{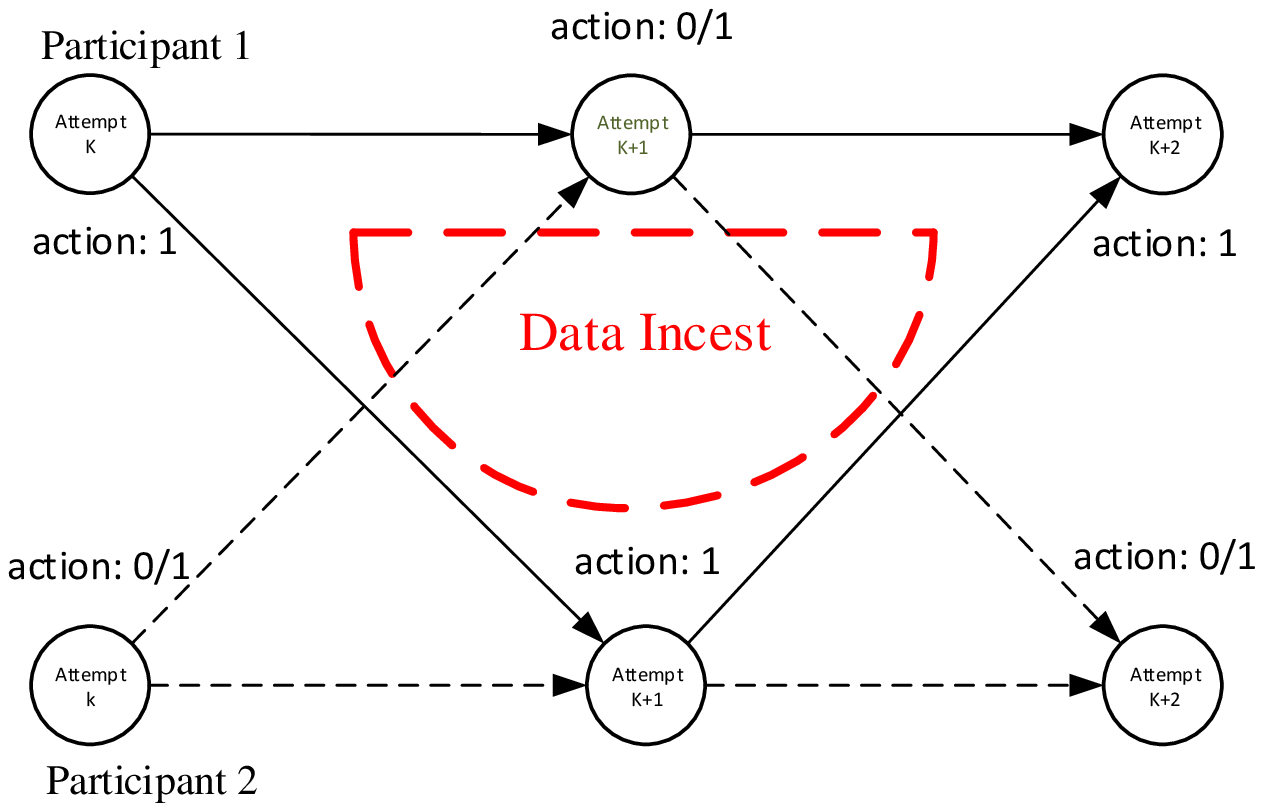}}
\subcaption{}
\label{diss}
\end{minipage}
\begin{minipage}[b]{.45\textwidth}
\hspace{.2cm}\scalebox{.57}{\includegraphics{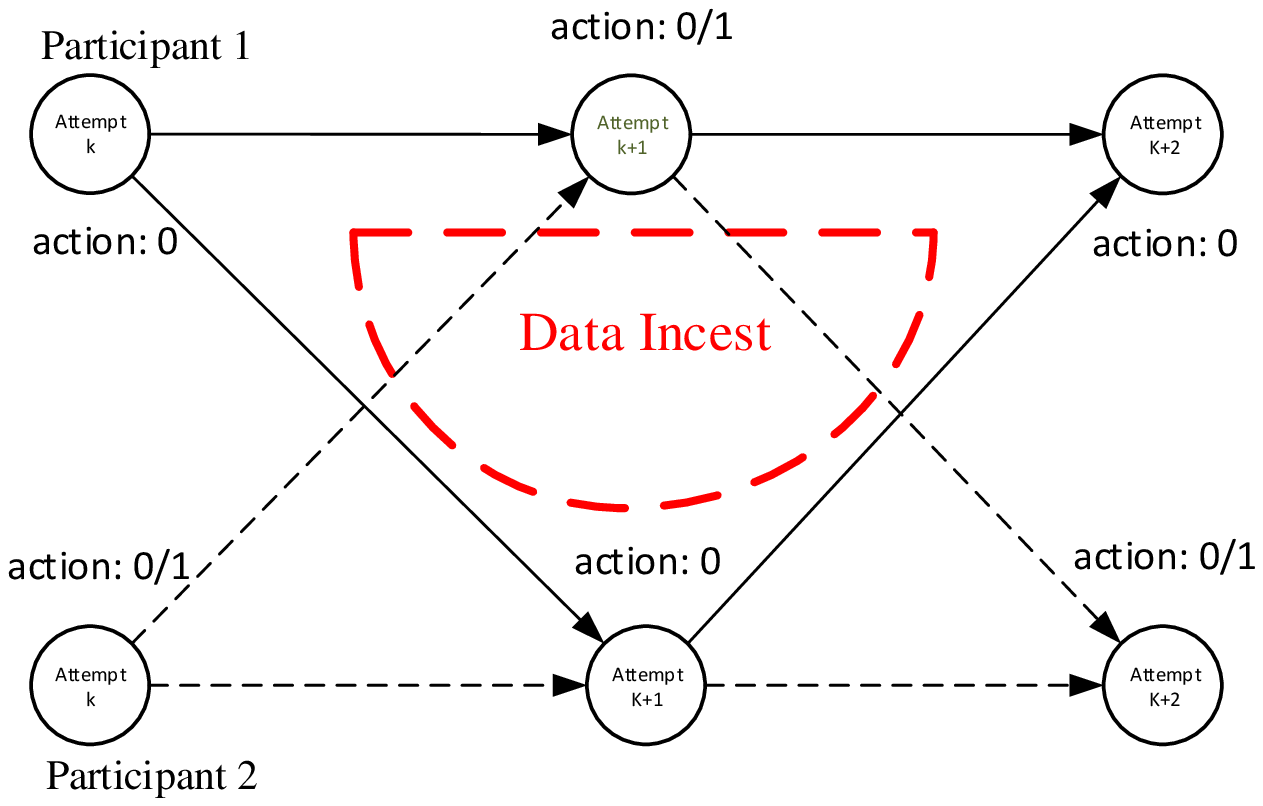}}
\subcaption{}
\label{diss2}
\end{minipage}
\caption{\label{exp:dataincest} Two scenarios where data incest arose in our experimental studies.  Each participant acted several times in a predetermined sequential order. In each step, the attempt of Participant~1 happened before that of Participant~2. The links in the above graphs show the information flow between participants in each scenario. }
\end{figure}
\begin{figure}[h]
\centerline{
\includegraphics[width=.8\textwidth]{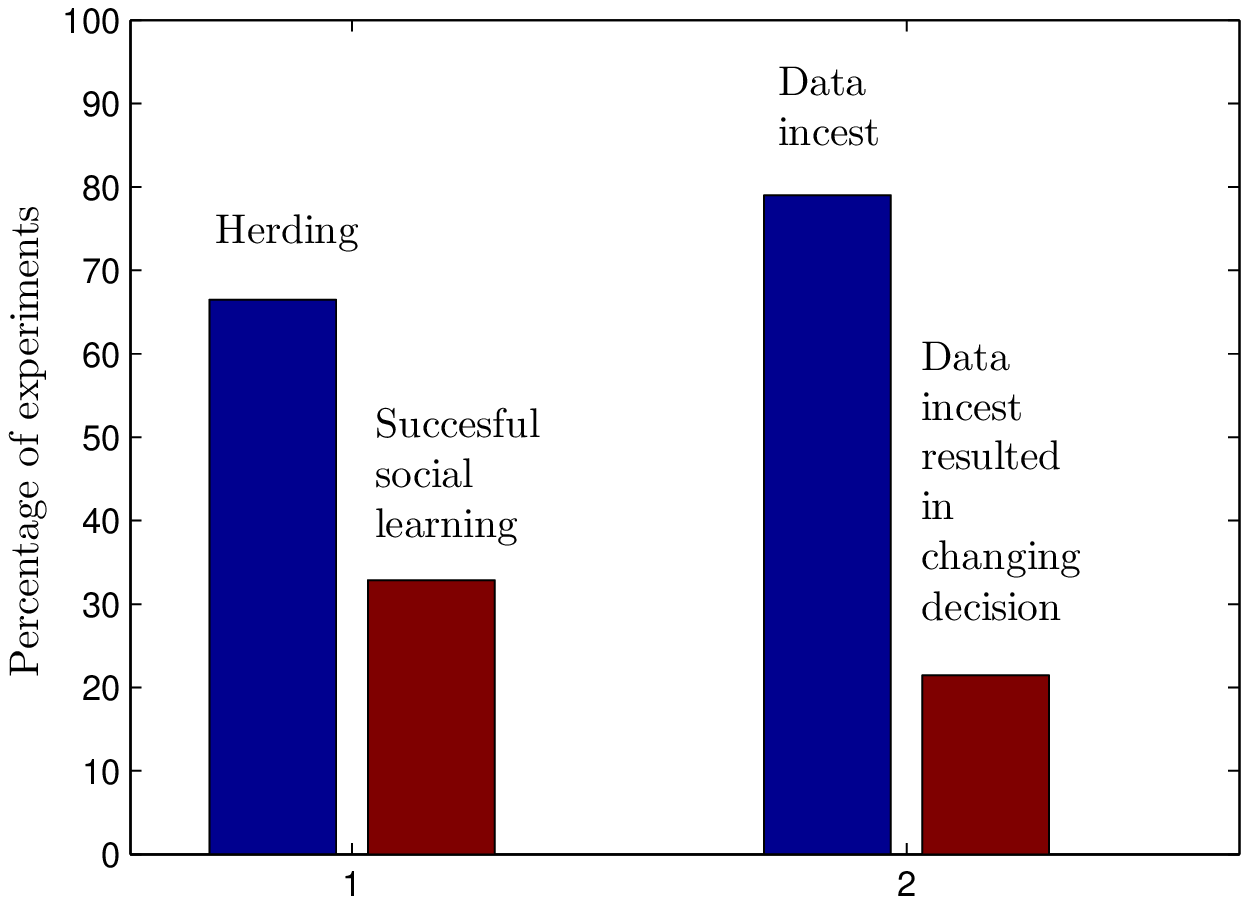}}
\caption{Social learning with data incest that exercised by groups of students who were asked to perform a conceptual task in our experimental study.}
\label{summary}
\end{figure}
\item \textit{Discussion:} Among 3316 (non-unique) participants of this experiment, 1336 observations (around 40\%) did not change their judgements after observing the action of their partners, while the other 60\% changed their initial judgment and were influenced by the action of their partners. An experimental observation that stands out here is that the individuals can be divided into two types: (i) \textit{boundary agents} who stand firm on their decisions during the trial, i.e., their decisions are independent of decisions of the other agents and (ii) \textit{internal agents} who are affected by decisions of the other agents.   Fig.~\ref{Fig:Samplepath} shows the sample path of two participants in a  group, Participant~1  is an internal node while Participant~2 is a boundary node.

 \begin{figure}[htb]
\centerline{
\includegraphics[width=.8\textwidth]{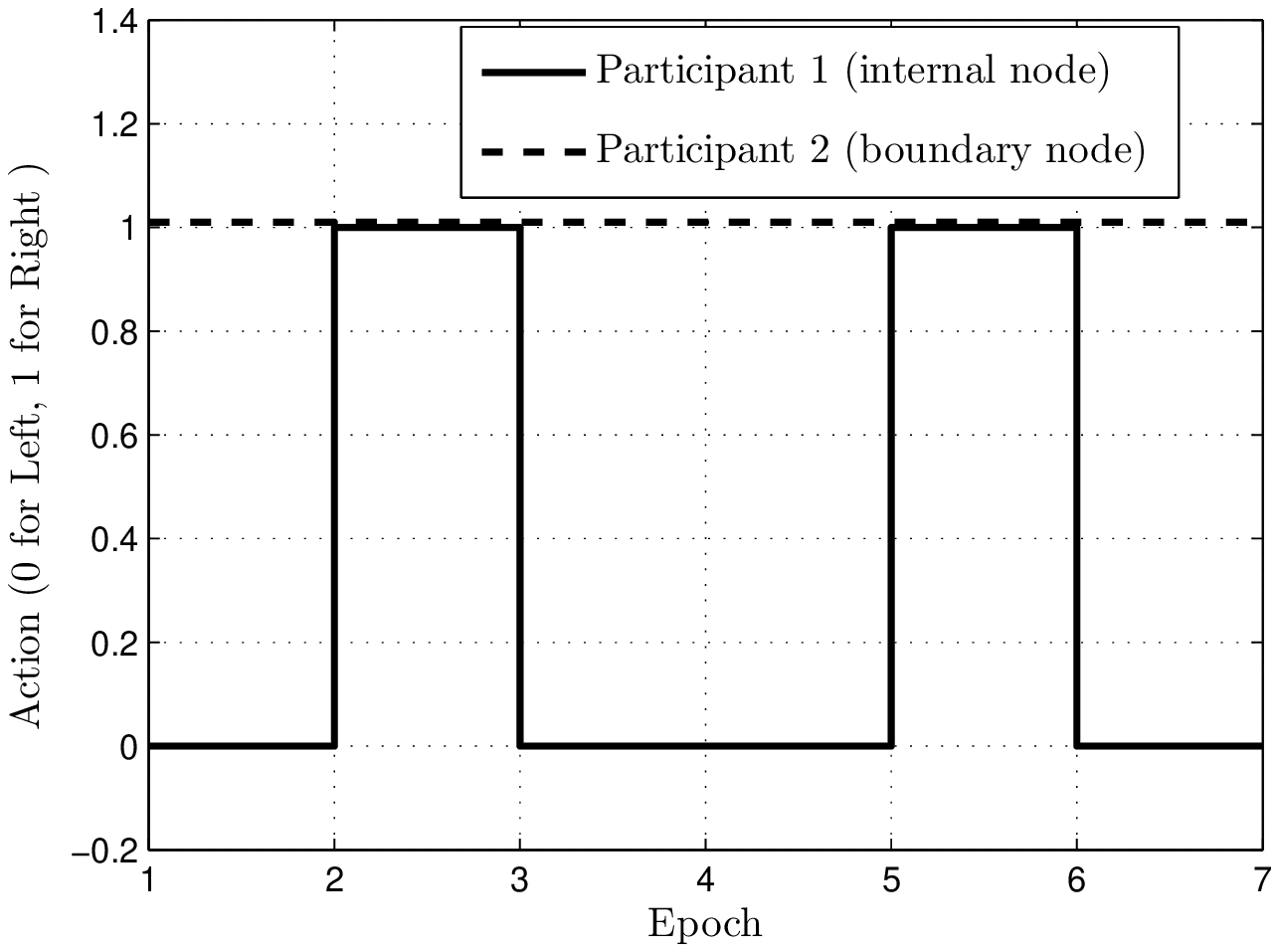}}
\caption{Actions of two participants in a group at different epochs. Participant~1 can be considered as an internal node and Participant~2 can be viewed as a boundary node.}
\label{Fig:Samplepath}
\end{figure}

 To study the decision making behavior of individuals of each type,  we investigate the time taken by each participant to make his judgment. Let $\mu_{\rm judg.}$ and $\sigma_{\rm judg.}$ denote the mean and the standard deviation of the time taken by participants to make their judgments in milliseconds. The results of our experimental study, which are presented in Table~I, show that the internal nodes,  on average, required more time to make their judgments compared to the boundary nodes; this is quite intuitive from the fact that the boundary nodes stood firm on their decisions and ignored the judgment of their partners and thus required less time to make their judgments.

\begin{table}[h]
\begin{center}
 \begin{tabular}{ | c || c | l | l |  }
 \hline
  Type of nodes & relative frequency & $\mu_{\rm judg.}$ & $\sigma_{\rm judg.}$ \\ \hline
  Internal & 40 \% & 1058 ms & 315 ms \\ \hline
  Boundary & 60 \% & 861 ms & 403 ms\\ \hline
\end{tabular}
 \end{center}
 \label{table1}
 \caption{The frequency of the internal and the boundary nodes in a community of 3316 undergraduate students of the University of British Columbia along with the statistics of the time required by participants (of both types) to make their judgments in milliseconds.}
 \end{table}
 \end{itemize}
\section{Conclusion}
We conducted a psychology experiment on a group of undergraduate students to study the learning behavior of humans in a society. The parameters of a social learning model (cost function, observation probabilities, and initial probability distribution of  state of nature) were fitted to the actions of participants of the psychology experiment. Our experimental study shows that the individuals who stand firm on their decisions require less time to make their decision (on average) compared to those who learn from decisions of other agents. We further  showed that, as a result of the information exchange protocol (between individuals within a group) and the recursive nature of decision making process, data incest arises in a  large fraction of trials in the experiment.
\bibliographystyle{IEEEtran}
\bibliography{ref_Maziyar,ref_Vikram}
\end{document}